\documentclass{PoS}

\PoS{PoS(LAT2005)334}

\title{The Gluon Propagator in Lattice Landau Gauge with twisted 
	boundary conditions }

\ShortTitle{The Gluon Propagator in Lattice Landau Gauge with twisted 
	boundary conditions }

\usepackage{epsfig}
\usepackage[all]{xy}
\usepackage{amsfonts,bbm}
\usepackage{amsmath}
\usepackage{eucal}
\usepackage{amssymb}
\usepackage{graphics}

\newcommand{\Id}{\mathbbm{1}}

\author{\speaker{Torsten Tok}\thanks{Supported by the {\em
       Deutsche Forschungsgemeinschaft} (DFG), contracts
       SM 70/1-2 and Re 856/4-3.}\\

       Institut f\"ur Theoretische Physik, Universit\"at T\"ubingen,
       D-72076 T\"ubingen, Germany\\
       Institut f\"ur Theoretische Physik III, Universit\"at
       Erlangen, D-91058, Erlangen, Germany

       E-mail: \email{tok@tphys.physik.uni-tuebingen.de}}

\author{Kurt Langfeld, Hugo Reinhardt\\

       Institut f\"ur Theoretische Physik, Universit\"at T\"ubingen,
       D-72076 T\"ubingen, Germany}

\author{Lorenz von Smekal\\

       Centre for the Subatomic Structure of Matter, School of
       Chemistry \& Physics,\\ The University of Adelaide,
       Adelaide, SA 5005, Australia}

\abstract{We investigate the infrared behaviour of the gluon 
propagator in Landau
gauge on a lattice with twisted boundary conditions. 

Analytic calculations using Dyson-Schwinger equations, exact
renormalization group and stochastic quantization show that the 
gluon propagator in Landau gauge approaches zero for small momentum. 
On the other
hand lattice calculations and calculations on a four-torus 
seem to give rise to a non-zero limit. One
possible reason for this difference is the existence of 
zero-momentum fluctuation modes which potentially give a massive 
contribution to the gluon propagator. Our simulations show that with 
twisted boundary conditions these zero-momentum modes are suppressed and
the gluon propagator becomes smaller than in a periodic ensemble.
}

\FullConference{XXIIIrd International Symposium on Lattice Field Theory\\
		 25-30 July 2005\\
		 Trinity College, Dublin, Ireland}

\begin{document}

\section{Introduction}

On $\mathbbm{R}^4$ the gluon propagator has been calculated with several
techniques, using Dyson-Schwinger equations \cite{Lerche:2002ep}, 
stochastic quantization \cite{Zwanziger:2001kw} and exact renormalization 
group equation \cite{Pawlowski:2003hq}. Representing the gluon propagator 
in Landau gauge as 
\begin{equation}
D_{\mu \nu}^{a b} (p) = 
\delta^{a b} (\delta_{\mu \nu} - p_\mu p_\nu / p^2) D ( p^2 )
\end{equation}
the low momentum behaviour was found to be 
$D(p^2) \stackrel{p \to 0}{\propto} (p^2)^{(2 \kappa -1)}$ with 
$\kappa \approx 0.6$. On the other hand calculations on the torus 
$\mathbbm{T}^4$ using Dyson-Schwinger equations \cite{Fischer:2002hn} 
and lattice
calculations \cite{Gattnar:2004bf,Sternbeck:2005tk,%
Bonnet:2001uh,Bowman:2004jm,Oliveira:2004gy} 
seem to indicate a non-zero value of $D$ at $p=0$, 
i.e.~$\kappa = 0.5$. A possible explanation for the difference 
between $\mathbbm{R}^4$ (non-compact space-time) and 
$\mathbbm{T}^4$ (compact space-time) are zero 
momentum modes on the compact space-time manifold which potentially 
give rise to a massive contribution to the gluon propagator. 

In this talk we wish to investigate how the implementation of non-trivial
boundary conditions changes the infrared behaviour of the gluon 
propagator.

\section{Twisted boundary conditions}

To determine the influence of different boundary conditions on the
gluon propagator on the lattice (or on the torus) we will consider
{\em twisted} boundary conditions. 

As usual we denote link variables by $U_\mu(x)$, the unit vector in
$\mu$-direction by $\hat \mu$ and the extension of the lattice in
$\mu$-direction by $L_\mu$. Gauge fixing can be implemented by
maximizing some functional $F[g,U]$ with respect to gauge
transformations 
$U_\mu(x) \to U^g_\mu(x) = g^{-1} (x) U_\mu(x) g(x+\hat \mu)$, e.g.~for
Landau gauge  we have $F[g,U] = \sum_{x,\mu} U_\mu^g(x)$. 
Non-trivial boundary conditions on the lattice (as well as on the torus) 
can be introduced by demanding that the link variable is periodic up to
a gauge transformation, i.e.
\begin{equation}
\label{periodicity}
U_\mu (x + L_\nu \hat \nu) = U_\mu^{\Omega_\nu} (x) =
\Omega_\nu^{-1} (x) U_\mu (x) \Omega_\nu (x+\hat \mu) \, ,
\end{equation}
where $\Omega_\nu (x)$ is called transition function. This property of
the link variable ensures that gauge invariant observables are actually 
periodic. The transition functions have to fulfill the cocycle condition
\begin{equation}
\Omega_\nu (x) \Omega_\mu (x + L_\nu \hat \nu) = 
\Omega_\mu (x) \Omega_\nu (x + L_\mu \hat \mu) Z_{\mu \nu} \, , 
\end{equation}
where $Z_{\mu \nu}$ are elements of the center of the gauge group. 
Twisted boundary conditions imply that some of the group elements 
$Z_{\mu \nu}$ are non-trivial. Under a general gauge transformation 
$g$ the transition functions transform as 
\begin{equation}
\label{trans_fct_trans}
\Omega^g_\mu (x) =  g^{-1} (x) \Omega_\mu (x) g (x + L_\mu \hat \mu)
\end{equation}
leaving the twists $Z_{\mu \nu}$ unchanged. For twisted boundary
conditions gauge fixing is non-trivial: If we implement non-trivial 
boundary conditions via some transition functions $\Omega_\mu (x)$ 
we should keep these transition functions fixed during gauge fixing, 
i.e.~we have to find the maximum of the gauge fixing functional 
$F[g,U]$ with respect to such gauge transformations $g$ which do not 
change the transition functions. 

But there is a second problem concerning the choice of
transition functions for a given gauge functional $F[g,U]$. Let us
consider the Landau gauge in periodic boundary conditions. Maximizing 
$F[g,U]$ in this case means that we try to make the deviation of the
gauge fixed configuration $U^g_\mu(x)$ from the configuration 
$U_\mu(x) \equiv \Id$ (which corresponds to the absolute maximum of $F$ 
and which has zero field strength) as small as possible. On the other 
hand one could (awkwardly) choose 
non-trivial transition functions\footnote{This can be achieved by a 
{\em non-periodic} gauge transformation $g$. Then, according to
eq.~(\ref{trans_fct_trans}), one has the transition functions
$\Omega_\mu(x) = g^{-1}(x) g (x + L_\mu \hat \mu) \neq \Id$.} 
and keep them fixed during Landau gauge fixing. Then in general 
$U_\mu(x) \equiv \Id$ will not fulfill these new periodicity 
properties (\ref{periodicity}) and some other  
configuration $\tilde U_\mu(x)$ will be the maximum of the functional 
$F[\Id,U]$. However, in general this configuration $\tilde U_\mu(x)$ 
will have non-zero and non-constant field strength. 
Landau gauge fixing with these (artificially) introduced
transition functions will make the deviation of the
gauge fixed configuration $U^g_\mu(x)$ from $\tilde U_\mu(x)$ as small
as possible. This will obviously destroy translational invariance, which
is physically not sensible. The lesson is that one should be careful 
in choosing the transition functions. 
If possible one should choose the transition functions (also in
the twisted case) such  that $U_\mu(x) \equiv \Id$ fulfills 
eq.~(\ref{periodicity}).

From now on we will specialize to the gauge group $SU(2)$. 
We parametrize the links as usual ($\sigma_k$ - Pauli matrices)
\begin{equation}
U_\mu (x) = u_\mu^0 (x) \Id + i u_\mu^k (x) \sigma_k \, , \quad 
u_\mu^j \in \mathbbm{R} \, , \quad j = 0,1,2,3 
\end{equation}
and choose constant transition functions, so-called twist eaters
\cite{Ambjorn:1980sm}:
\begin{equation}
\Omega_0 \equiv \Id \, , \quad 
\Omega_k \equiv i \sigma_k \, , \quad k = 1,2,3
\end{equation}
corresponding to the twists
\begin{equation}
Z_{0 k} = \Id \, , \quad k = 1,2,3 \, , \quad 
Z_{1 2} = Z_{2 3} = Z_{3 1} = - \Id \, .
\end{equation}
From eq.~(\ref{periodicity}) one easily obtains
\begin{equation}
u_\mu^0 (x + L_k \hat k) =  u_\mu^0 (x) \, , \quad 
u^a_\mu (x + L_k) = \left\{ \begin{array}{ll}
			u^a_\mu (x) \, , \, & {\mathrm{if}} \quad  k=a \\
			- u^a_\mu (x) \, , \, & {\mathrm{else}}
			\end{array} \right. \, ,
\end{equation}
i.e.~some of the colour components are anti-periodic in spatial directions.
The vacuum configuration $U_\mu (x) \equiv \Id$ 
fulfills these boundary conditions, which are compatible with Landau
gauge.

\begin{figure}[!t]
  \centerline{
        \epsfig{file=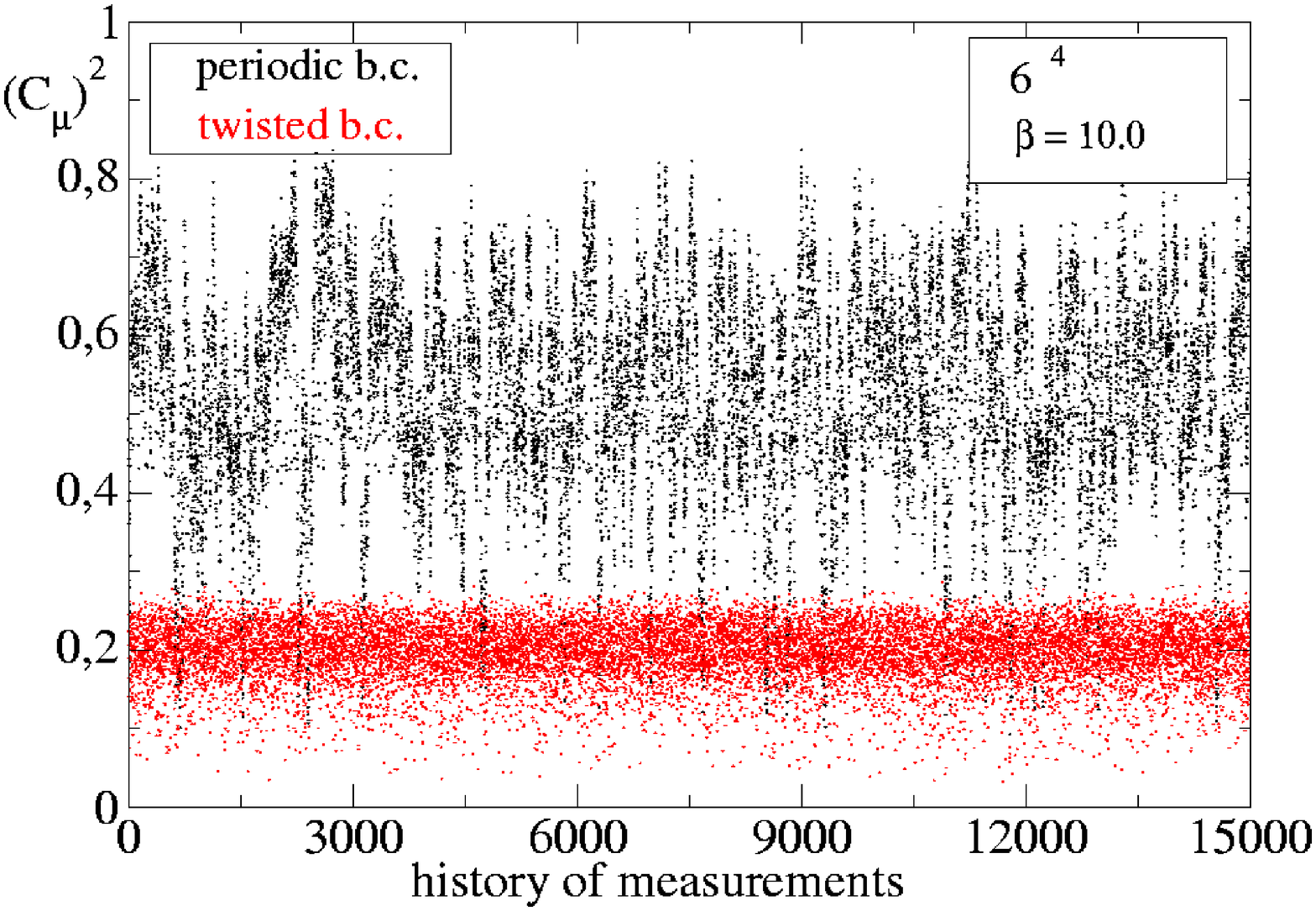,width=7cm,height=6cm}
        \hfill
	\epsfig{file=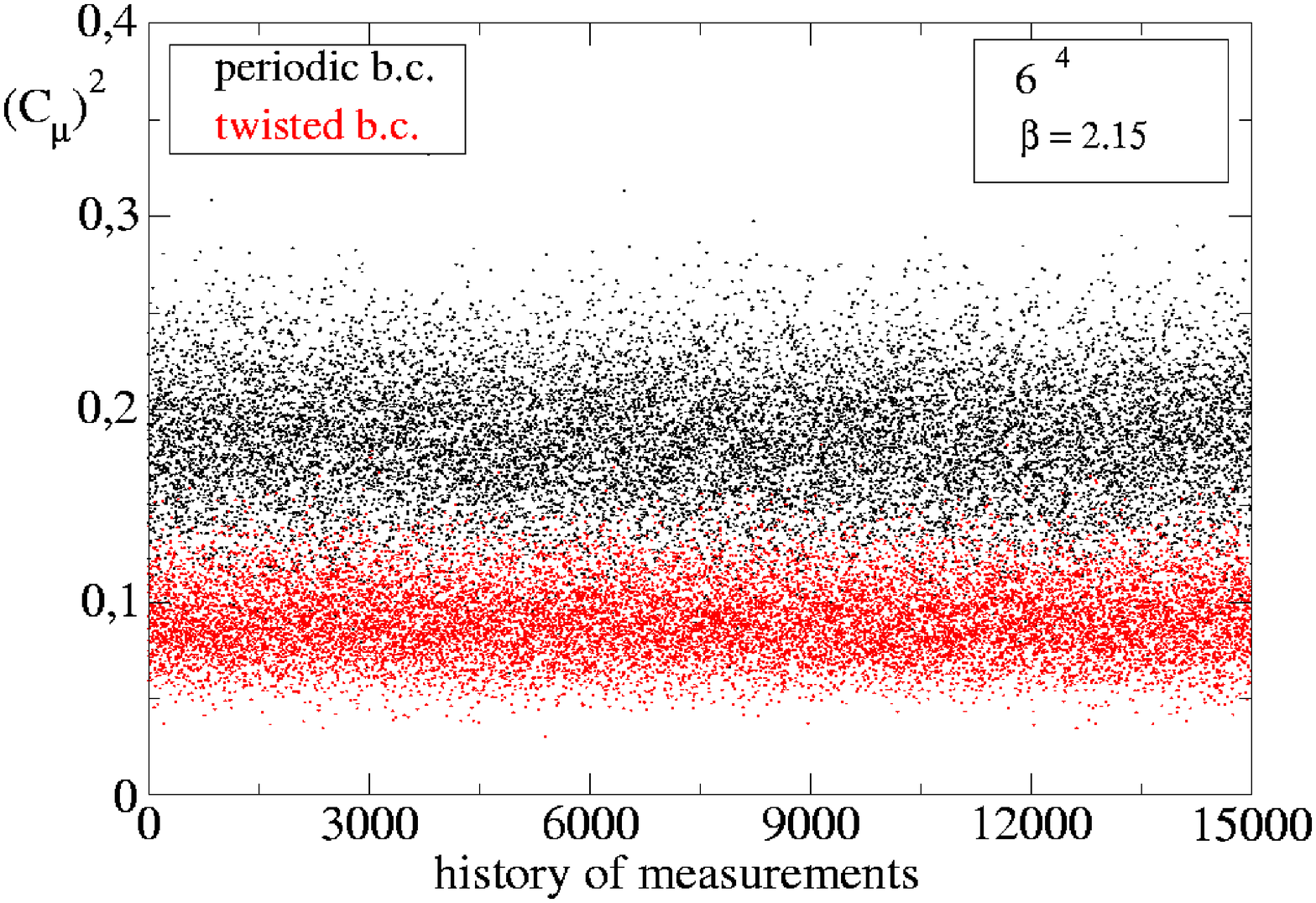,width=7cm,height=6cm}}
        \caption{Simulation history of the zero momentum part $C_\mu$ in
	periodic (black) and twisted (red) boundary conditions on a 
	$6^4$ lattice with $\beta=10$ (left) and $\beta=2.15$ (right). 
	\label{fig:01}}
\end{figure}

Now let us have a look at the zero momentum part of the gauge potential.
The gauge potential $A_\mu (x)$ (corresponding to the links $U_\mu(x)$) 
and and its Fourier transformation are given by
\begin{eqnarray} 
A_\mu (x) &=& 1/2 (U_\mu(x) - U_\mu^+ (x)) \\
\tilde A_\mu (p) &=& \frac{1}{V} \sum_{x} A_\mu (x) e^{i p x} 
\end{eqnarray}
With periodic boundary conditions the zero-momentum part 
$C_\mu := \tilde A_\mu (p = 0)$ shows large fluctuations during 
the simulation \cite{Damm:1998pd}, as can be seen in fig.~\ref{fig:01}.
These fluctuations and their amplitudes are clearly reduced with 
twisted boundary conditions.
At this point let us mention two further advantages of the constant
transition functions: They are numerically easy to implement and the 
periodicity in time allows finite temperature calculations. 

The effect of the twisted boundary conditions should become smaller
on larger lattices, while for small lattices we expect a strong
dependence on the lattice size.

\section{The gluon propagator in twisted boundary conditions}

\begin{figure}[!t]
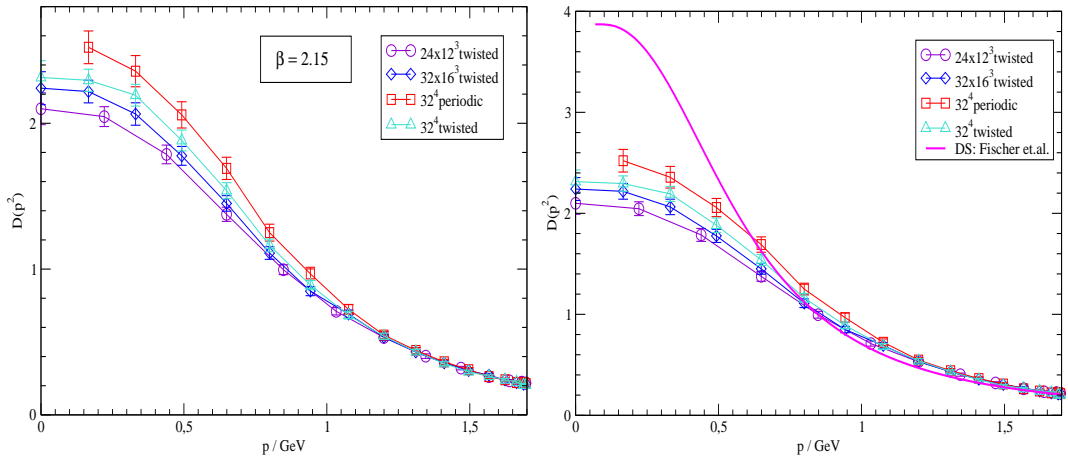

  \textsf{
  \begin{center}
        \epsfig{file=pics/wir_7.eps,width=7cm,height=6cm}
        \epsfig{file=pics/christ_wir_1.eps,width=7cm,height=6cm}
        \caption{Plot of gluon propagator $D(p^2)$ with twisted and
	periodic boundary conditions for different lattice sizes (left)
	and comparison with results from Dyson-Schwinger equation on the
	Torus \cite{Fischer:2002hn}.
	\label{fig:02}}
  \end{center}
  }
\end{figure}

We have measured the gluon propagator for different
lattice sizes at $\beta = 2.15$, see fig.~\ref{fig:02}. 
As can be seen: for larger spatial extension of the 
lattice the ``twisted'' gluon propagator becomes larger and the
``twisted'' gluon propagator is always smaller than the ``periodic'' gluon 
propagator, i.e.~one can consider the gluon propagator 
in the twisted ensemble as a lower bound for the gluon propagator in 
the periodic ensemble. As expected for larger spatial extensions the 
difference between the propagators with periodic and twisted boundary
conditions decreases, i.e.~the finite size effects due to the twists 
become smaller. 

All these observations suggest that the gluon 
propagator $D(p^2)$ on the lattice has a non-zero limit for $p \to 0$.
Furthermore, our results confirm results obtained by solving Dyson-Schwinger
equations on the torus \cite{Fischer:2002hn}.

Unfortunately, our investigations do not clarify the difference between
compact and non-compact space-times. Our findings seem to indicate that
this difference in the infrared behaviour of the gluon propagator
remains even in the limit of an infinitely large lattice.

\section*{Acknowledgement}

We would like to thank the organizers of Lattice 2005 for the
organization of this interesting conference. 

\end{document}